\documentclass[prd,superscriptaddress,amsfonts,amssymb,amsmath,showpacs,twocolumn]{revtex4-2}
\usepackage{bm}
\usepackage{amsfonts}
\usepackage{latexsym}
\usepackage[latin1]{inputenc}
\usepackage{graphicx}
\usepackage{amsmath}
\usepackage{palatino}
\usepackage{mathpazo}
\usepackage{textcomp}
\usepackage{float}
\usepackage{booktabs}
\usepackage{dcolumn}
\usepackage{ragged2e}
\usepackage{hyperref}
\hypersetup{colorlinks,citecolor=blue}
\hypersetup{colorlinks=true,linkcolor=red,filecolor=magenta,    urlcolor=blue}
\usepackage{amsmath}
\usepackage{xcolor}
\usepackage{orcidlink}
\usepackage[caption=false]{subfig}
\usepackage{commath}
\def\jnl@style{}
\def\aaref@jnl#1{{\jnl@style#1}}

\def\aaref@jnl#1{{\jnl@style#1}}

\def\aj{\aaref@jnl{AJ}}                   
\def\apj{\aaref@jnl{ApJ}}                 
\def\apjl{\aaref@jnl{ApJ}}                
\def\apjs{\aaref@jnl{ApJS}}               
\def\apss{\aaref@jnl{Ap\&SS}}             
\def\aap{\aaref@jnl{A\&A}}                
\def\aapr{\aaref@jnl{A\&A~Rev.}}          
\def\aaps{\aaref@jnl{A\&AS}}              
\def\mnras{\aaref@jnl{Mon.~Not.~Roy.~Astron.~Soc.}}             
\def\prd{\aaref@jnl{Phys.~Rev.~D}}        
\def\plb{\aaref@jnl{Phys.~Lett.~B}}        
\def\prc{\aaref@jnl{Phys.~Rev.~C}}  
\def\prl{\aaref@jnl{Phys.~Rev.~Lett.}}    
\def\qjras{\aaref@jnl{QJRAS}}             
\def\skytel{\aaref@jnl{S\&T}}             
\def\ssr{\aaref@jnl{Space~Sci.~Rev.}}     
\def\zap{\aaref@jnl{ZAp}}                 
\def\nat{\aaref@jnl{Nature}}              
\def\aplett{\aaref@jnl{Astrophys.~Lett.}} 
\def\apspr{\aaref@jnl{Astrophys.~Space~Phys.~Res.}} 
\def\physrep{\aaref@jnl{Phys.~Rep.}}      
\def\physscr{\aaref@jnl{Phys.~Scr}}       
\def\commat{\aaref@jnl{Comm.~Math.~Phys.}}              
\def\science{\aaref@jnl{Science}}               
\def\cqg{\aaref@jnl{Classical Quant.~Grav.}}            
\def\jpcs{\aaref@jnl{JPCS}}                                     
\def\ijmpd{\aaref@jnl{Int.~J.~Mod.~Phys.~D}}                    
\def\grg{\aaref@jnl{Gen.~Relat.~Gravit.}}               
\def\rpp{\aaref@jnl{Rep.~Prog.~Phys.}}          
\def\npa{\aaref@jnl{Nucl.~Phys.~A}}        
\def\lrr{\aaref@jnl{Living Rev.~Rel.}}                   
\def\jcap{\aaref@jnl{J.~Cosmology Astropart.~Phys.}}    
\def\rmp{\aaref@jnl{Rev.~Mod.~Phys.}}   
\def\epjc{\aaref@jnl{Eur.~Phys.~J.~C}}

\allowdisplaybreaks[1]

\begin{document}
\color{black}       

\title{Theory of gravity with nonminimal matter-nonmetricity coupling and the de-Sitter swampland conjectures}

\author{Sanjay Mandal\orcidlink{0000-0003-2570-2335}}
\email{sanjaymandal960@gmail.com}
\affiliation{Faculty of Mathematics and Computer Science, Transilvania University, Iuliu Maniu Str. 50, 500091 Brasov, Romania.}

\author{Kazuharu Bamba\orcidlink{}}
\email{bamba@sss.fukushima-u.ac.jp}
\affiliation{Faculty of Symbiotic Systems Science, Fukushima University, Fukushima 960-1296, Japan.}

%
\date{\today}
\begin{abstract}
In this study, we investigate swampland conjectures within the setup of matter and non-metricity nonminimal coupling theories of gravity. We examine how the inflationary solution produced by a single scalar field can be resolved with the swampland criteria in string theory regarding the formation of de Sitter solutions. The new important findings are that the inflationary scenario in our study differs from the one in general relativity because of the presence of a nonminimal coupling term, and that difference gives the correction to general relativity. In addition, we observe that the slow-roll conditions and the swampland conjectures are incompatible with each other for a single scalar field within the framework of nonminimally coupled alternative gravity theories. We predict that these results will hold for a wide range of inflationary scenarios in the context of nonminimal coupling gravitational theories.\\


\end{abstract}

\maketitle

\section{Introduction}

The $\Lambda$CDM model, the prevailing framework in cosmology, effectively captures and explains contemporary cosmological and astrophysical observations. It relies on two key components that extend beyond the standard model of particle physics: dark matter and dark energy. Dark matter is postulated as a particle within theories surpassing the standard model, while dark energy remains a puzzle lacking a fully satisfactory solution. The cosmological constant emerges as the most straightforward candidate for dark energy, and observational data align with its predictions. According to this model, the distant future of the Universe involves an asymptotic transition to de Sitter spacetime, characterized by a constant expansion rate. Despite ongoing efforts to reconcile particle physics and general relativity, there is still no definitive explanation within current theories for the origin of the cosmological constant, such as residual vacuum energy density. Recent discussions question the feasibility of realizing de Sitter spacetime in string theory, as highlighted in references \cite{sw1,sw2,sw3}, with \cite{In4} offering a comprehensive review and \cite{5a} sounding a note of caution regarding the swampland conjectures.

The swampland conjectures are a series regarding the circumstances that must be satisfied in order to accept local gauge symmetries and at least one Planck mass particle in order to account for gravity's weakness. Superluminal propagation must additionally avoid high-order elements in the effective action (for a discussion, see Ref. \cite{In4}). To the best of our knowledge, no assumptions are involved in the Strong Equivalence Principle in this scenario, and assuming that General Relativity theory remains an inevitable gravitational theory. As a result, it is natural to investigate if the swampland conjectures hold true for alternative gravity theories that allow for single-field inflation. This applies to nonminimal matter and curvature coupling gravitational theories of gravity  \cite{In5, In6}, which might provide inflationary solutions.

String gravitational theories are more complicated than general relativity. But, the effective string theory models demonstrate extra terms in the form of coupling between high-order curvature terms and derivatives of the dilaton field; those extra terms represent the adjustments to general relativity \cite{In7, In8, In9, In10, In11}.

However, alternative gravity theories are proposed to search for possible avenues for exploring astrophysical and cosmological scenarios, for example, the flattening of the rotation curves of galaxies and the accelerated expansion of the Universe, instead of resorting to dark matter and dark energy, independent of string and quantum gravity considerations. Modified theories of gravity not only successfully address the late-time cosmologies but are also able to discuss the early-time scenario of the universe. For instance, one can see the references herein for the detail studies in modified theories of gravity \cite{Nojiri:2006ri, Copeland:2006wr, Durrer:2007re,Bamba:2012cp,Sotiriou:2008rp, DeFelice:2010aj,Nojiri:2010wj, Capozziello:2011et,Clifton:2011jh,Joyce:2014kja,Cai:2015emx,Nojiri:2017ncd,Langlois:2018dxi,Frusciante:2019xia,Bahamonde:2021gfp,Arai:2022ilw,Odintsov:2023weg}. Further, the $\Lambda$CDM model predicts that the universe will approach de Sitter spacetime with a constant speed in the far future, whereas it is argued that the de Sitter space-time cannot be realized in string theory \cite{In4}. The implications of swampland conjectures for dark energy or modified gravity with the de Sitter solutions investigated widely in Literature \cite{Heisenberg:2018yae, Agrawal:2018own,DavidMarsh:2018etu,vandeBruck:2019vzd,Brax:2019rwf, Benetti:2019smr}. In \cite{Heisenberg:2018yae}, the authors have examined the implications of swampland conjectures on dark energy and strictly constrain the standard dark energy model's parameter. The bounds on the model parameter are in agreement with the string swampland criteria. Also, they indicated that future investigation is needed to measure the tension between quintessence models and the string Swampland criteria. Agrawal et al. \cite{Agrawal:2018own} presented bounds on the scalar potential and argued that the universe goes under a phase transition within a few Hubble time. Bruck and Thomas presented an interesting investigation in swampland criteria, in which they argued that the coupling between scalar field and dark matter helps us to reduce the tension between swampland conjectures and the properties of quintessential potential \cite{vandeBruck:2019vzd}. Moreover, the swampland criteria are applied to solve the important issues in modern observational cosmology, i.e., $H_0$ tensions due to the coupling between scalar field and dark matter\cite{OColgain:2018czj,Agrawal:2019dlm, Colgain:2019joh}. So far, we have discussed the studies examined at the basic level of gravitational theories. As we know, the alternative modified gravitational theories are very successful in presenting the dark energy, dark matter, and late-time acceleration issues; therefore, investigating swampland criteria in the context of a modified gravity scenario will open many ways to explore the above-discussed issues. 

Moreover, a well-established gravitational theory, known as $f(Q)$ theory \cite{In12}, where the non-metricity scalar $Q$ is replaced by a generic function of $f(Q)$ in Einstein-Hilbert action. Further, a possible generalization of symmetric teleparallel equivalent to general relativity is to non-minimal coupling between matter and non-metricity by introducing two general functions of $Q$ as $f_1(Q)$ and $f_2(Q)$, in the Einstein-Hilbert action \cite{Harko/2018}. The function $f_1(Q)$ has a similar role as $f(Q)$ gravity, whereas the $f_2(Q)$ multiplies with the matter Lagrangian ($\mathcal{L}_m$) gives rise to the coupling between matter and non-metricity. In literature, the $f(Q)$ gravitational theories have been studied extensively (for instance, one can see a review article on $f(Q)$ gravity \cite{In14}), whereas the nonminimal coupling $f(Q)$ theory has not been studied widely after its proposal. Thereafter, only one article was published by Mandal and Sahoo to explore the accelerated expansion scenario using the observational constraint on the equation of state parameter ($\omega$) in the context of nonminimal coupled theory \cite{In15}. Therefore, we aim to explore swampland criteria in this theory and present one of the many possible applications of this coupled theory. We are the first to explore this type of investigation, which will be extended to many more applications related to this study.

In this study, we shall consider the nonminimally coupled matter and non-metricity framework with a single inflationary field system to be discussed in the upcoming sections. Our presentation starts with a detailed discussion of the non-minimal coupled theory framework in section \ref{sec2}. Thereafter, we present the inflationary scenario associated with the nonminimal coupled theory in section \ref{sec3}. Also, we will show that in the nonminimal coupled model, regardless of the similarity among the warm inflation and slow-roll parameters, the swampland conjectures cannot be fulfilled in the former case. In last, we conclude our results in section \ref{sec5}.

\section{Non-minimal matter-nonmetricity coupling theory of gravity}\label{sec2}

Let us start by considering the action for non-minimal matter-nonmetricity coupling theory of gravity proposed by Harko et al., \cite{Harko/2018}
\begin{equation}
\label{1}
S=\int d^4x \sqrt{-g}\left[\frac{M_p^2}{2}f_1(Q)+f_2(Q)\mathcal{L}_m\right],
\end{equation}
here, $f_1(Q) \,\& \, f_2(Q)$ are the arbitrary Lagrangian functions of the non-metricity scalar $Q$, $g$ is the determinant of metric, and $\mathcal{L}_m$ is the matter Lagrangian.

Before proceeding further for further setup, let us present some of the nonmetricity tensors, and its traces are given below
\begin{equation}
\label{2}
Q_{\gamma\mu\nu}=\nabla_{\gamma}g_{\mu\nu}\,,
\end{equation}
\begin{equation}
\label{3}
Q_{\gamma}={{Q_{\gamma}}^{\mu}}_{\mu}\,, \qquad \widetilde{Q}_{\gamma}={Q^{\mu}}_{\gamma\mu}\,.
\end{equation}
In addition, the superpotential as a function of the nonmetricity tensor is given by
\begin{equation}
\label{4}
4{P^{\gamma}}_{\mu\nu}=-{Q^{\gamma}}_{\mu\nu}+2Q_{({\mu^{^{\gamma}}}{\nu})}-Q^{\gamma}g_{\mu\nu}-\widetilde{Q}^{\gamma}g_{\mu\nu}-\delta^{\gamma}_{{(\gamma^{^{Q}}}\nu)}\,,
\end{equation}
and, the nonmetricity scalar $Q$, which is the trace of non-metricity tensor has the following form, given by\cite{Harko/2018}
\begin{equation}
\label{5}
Q=-Q_{\gamma\mu\nu}P^{\gamma\mu\nu}\,.
\end{equation}
To simplify the formulation, let us introduce the following notations
\begin{align}
\label{6}
f=M_p^2\,f_1(Q)\,+2 f_2(Q)\mathcal{L}_m,\\
F=M_p^2\,f_1'(Q)+2 f_2'(Q)\mathcal{L}_m,
\end{align}
where primes (') represent the derivatives of functions $f_1(Q) \,\& \, f_2(Q)$ with respect to $Q$.

By definition, the energy-momentum tensor for the matter is essential for deriving the field equations, which can be written as,
\begin{equation}
\label{6}
T_{\mu\nu}=-\frac{2}{\sqrt{-g}}\frac{\delta(\sqrt{-g}\mathcal{L}_m)}{\delta g^{\mu\nu}}\,.
\end{equation}
Taking the variation of action \eqref{1} with respect to metric tensor, one can find the gravitational field equations given by
\begin{multline}
\label{7}
\frac{2}{\sqrt{-g}}\nabla_{\gamma}\left( \sqrt{-g}F {P^{\gamma}}_{\mu\nu}\right)+\frac{M_p^2}{2}g_{\mu\nu}f_1 \\
+F \left(P_{\mu\gamma i}{Q_{\nu}}^{\gamma i}-2Q_{\gamma i \mu}{P^{\gamma i}}_{\nu} \right)=-f_2 T_{\mu\nu}\,.
\end{multline}

In addition, the variation of the action with connection arises the following relation for Hyper-momentum tensor density
\begin{equation}
    H_{\lambda}^{\mu\nu}=-\frac{1}{2}\frac{\delta(\sqrt{-g}\mathcal{L}_m)}{\delta \Gamma^{\lambda}_{\mu\nu}}\,.
\end{equation}
Also, we can obtain
\begin{equation}
    \nabla_{\mu}\nabla_{\nu}\left( \sqrt{-g}F P^{\mu\nu}_{\alpha}-f_2 H_{\alpha}^{\mu\nu}\right)=0
\end{equation}
The detailed study on the above two relations related to the connection $\Gamma$ are discussed in \cite{Harko/2018}. With the above setup, we can investigate various cosmological scenarios in this modified alternative gravitational theory of nonminimal matter-nonmetricity coupling.

\subsection{The FLRW cosmology}\label{sec3}

To investigate the cosmology in this nonminimal gravitational theory, we consider a spatially flat metric with homogeneous and isotropic in nature, given by

\begin{equation}
\label{8}
ds^2 = -N^2(t) dt^2 +a^2(t) \delta_{ij} dx^i dx^j,
\end{equation}
where $i,\,\, j$ ranges over $1,2,3$ and represents the space coordinates, $N(t)$ represents lapse function, and we can take $N=1$ at any time for the usual time reparametrization freedom. $\delta_{ij}$ is the Kronecker delta. The rate of change in expansion and dilation can be written as
\begin{equation}
\label{9}
H=\frac{\dot{a}}{a},\,\,\,\,\ T=\frac{\dot{N}}{N},
\end{equation}
respectively. For the FLRW metric, the non-metricity scalar $Q$ reads as 
\begin{equation}
    Q=6 (H/N)^2.
\end{equation}

Using the above relations in the gravitational dynamic equations \eqref{7} with the perfect fluid description, the generalization of two Friedman equations can be written as:
\begin{equation}
\label{10}
f_2 \rho =\frac{M_p^2\,f_1}{2}-6 F \frac{H^2}{N^2},
\end{equation}
\begin{equation}
\label{11}
-f_2 p = \frac{M_p^2\,f_1}{2}-\frac{2}{N^2}[(\dot{F}-FT)H+F(\dot{H}+3H^2)],
\end{equation}
respectively. It is easy to verify that, for $f_1=-Q$ and $f_2=1=-F$, the above Friedman equations reduce to GR with cosmological constant $\Lambda$ \cite{Harko/2018}. Now, one can derive the continuity equation for the matter field  from the above motion equations,
\begin{equation}
\label{12}
\dot{\rho} +3 H(\rho+p) = -\frac{6 f_2'H}{f_2N^2}(\dot{H}-HT)(\mathcal{L}_m+\rho).
\end{equation}
If one wishes to recover the standard continuity equation from \eqref{12}, it can achieve by inserting $\mathcal{L}_m=-\rho$ in \eqref{12} and the results will be
\begin{equation}
\label{13}
\dot{\rho} +3 H(\rho+p) =0.
\end{equation}
As we know, the above conservation equation is compatible with the isotropic and homogeneous background of the universe, and the details about this equation in non-minimal coupling $f(Q)$ gravity are discussed in\cite{Harko/2018}.

\section{Inflation in nonminimal coupled theory}\label{sec3}

Before going to work on the inflationary profiles within the framework of the nonminimal coupled theory of gravity, let us briefly discuss the relations between the slow-roll parameters (i.e., $\epsilon$ and $\eta$, which play major roles in the inflationary scenario) and the swampland conjectures. To be more specific, the swampland conjectures restrict a few limits on the scalar fields arising at low energy, namely $\phi$, given by \cite{sw1,sw2},

\begin{equation}\label{sc1}
    \frac{\Delta \phi}{M_p}< c_1,
\end{equation}

\begin{equation}\label{sc2}
    M_p \frac{|\partial_{\phi}V|}{V}> c_2,
\end{equation}
where $\Delta \phi$ is the variation range of the scalar field, $M_p= M_{pl}/\sqrt{8\pi}$ is the reduced planck mass, $ \partial_{\phi}V=\frac{\partial V}{\partial \phi}$, $V(\phi)$ is the scalar field potential, and $c_1, c_2$ are contacts with order one. In further argument, one should consider a more redefined condition \cite{sw3,sw4,sw5}
\begin{equation}\label{sc3}
    M_p^2 \frac{|\partial^2_{\phi \phi}V|}{V}< - c_3,
\end{equation}
here, $ \partial^2_{\phi \phi}V=\frac{\partial^2 V}{\partial \phi^2}$ and $c_3$ is a constant of order one.

Now, the above conditions can be easily compared with the slow-roll parameters in the single-field inflation for the scalar field ($\phi$) \cite{sw6},
\begin{equation}
    \epsilon= \frac{M_p}{2}\left(\frac{|\partial_{\phi}V|}{V}\right)^2,
\end{equation}
and,
\begin{equation}
    \eta= M_p^2 \frac{\partial^2_{\phi \phi}V}{V}.
\end{equation}
It is well known that the slow-roll parameter satisfies $\epsilon\ll 1$ and $|\eta| \ll 1$ during the inflation period, whereas  $\epsilon\sim 1$ and $|\eta| \sim 1$ at the end of inflation. The latest constraints on the slow-roll parameters are presented by using the CMB data \cite{sw6,In16}
\begin{equation}
    \epsilon < 0.0044,
\end{equation}
\begin{equation}
    \eta= -0.015\pm 0.006.
\end{equation}
From these results, one can clearly see that the swampland conjecture parameters do not match the requirements on $c_2$ and $c_3$.

The incompatibility applies to any number of scalar fields driving inflation as long as their kinetic energy terms are canonical. \cite{In16}. Swampland theories, on the other hand, can be reconciled with data with warm inflationary model's setup \cite{In17, In18} for one \cite{In19, In20} or more scalar fields \cite{In16}.

Now, to test the above results in the nonminimal coupling theory of gravity, we shall proceed with $N=1$. For this choice, the non-metricity scalar $Q$ and  dilation rate $T$ reduce to
\begin{equation}
\label{14}
Q=6H^2,\,\,\,\, T=0.
\end{equation}
Moreover, the modified Friedman equations \eqref{10} and \eqref{11} can be rewritten as
\begin{equation}
\label{15}
3H^2= \frac{f_2}{2 F}\left(-\rho+\frac{M_p^2\,f_1}{2f_2}\right),
\end{equation}
\begin{equation}
\label{16}
\dot{H}+3H^2+\frac{\dot{F}}{F}H = \frac{f_2}{2 F}\left(p+\frac{M_p^2\,f_1}{2f_2}\right).
\end{equation}
Now, in our hand, we have two dynamical motion equations with five unknown quantities such as $H,\,\, \rho,\,\, p,\,\, f_1,\,\, f_2$.

To proceed further, we presume the Lagrangian function $f_1,\,\, f_2$ as follows
\begin{equation}
f_1(Q)=-Q,\,\, f_2(Q)=1+\alpha \bigg(\frac{Q}{6 M_p^2}\bigg)^n,
\end{equation}
for simplicity, we consider $\gamma=\frac{\alpha}{(6 M_p^2)^n}$.
Now, the field equations read
\begin{equation}\label{20}
    3 H^2=\frac{2 (1+\gamma Q^n)\, \rho+ M_p^2\, Q}{4(M_p^2-2 \gamma n Q^{n-1}\, p)},
\end{equation}
\begin{equation}\label{21}
    \Dot{H}=\frac{(1+\gamma Q^n)(\rho+p)-\frac{d}{dt}[2(\gamma n Q^{n-1})p]}{2(M_p^2-2 \gamma n Q^{n-1}\, p)}.
\end{equation}
In the above equations, we have also presumed that Lagrangian density $\mathcal{L}_m=p$ with a homogeneous scalar field $\phi$, for which the energy density and pressure given by
\begin{equation}\label{rho}
    \rho=\frac{\dot{\phi}^2}{2}+V(\phi),
\end{equation}
\begin{equation}\label{pres}
    p=\frac{\dot{\phi}^2}{2}-V(\phi).
\end{equation}
Furthermore, we assume quasi-exponential inflation i.e., $V\gg \dot{\phi}^2$, which implying $\rho\simeq -p \simeq V$ (from \eqref{rho} and \eqref{pres}, one can verify this relation easily). Then, from equation \eqref{20}, we can find

\begin{equation}\label{h2}
    -3 H^2 M_p^2+\gamma \, 6^n \left(H^2\right)^n (2 n p+\rho )+\rho =0.
\end{equation}

The measure for the weakness of gravity at least one Planck mass particle scale under the swampland conjectures is a set of hypotheses regarding the assumptions necessary to accept local gauge symmetries. Further, the higher-order terms in the effective theory are necessary to avoid the superluminal propagation (check a detailed study \cite{In4}). Therefore, we are proceeding with $n=3$. We would like to note here that for the choice of $n=1,2$, one can have similar types of results as for the choice of $n=3$.  And for this choice $n=3$, we find the solution for $H^2$, which is presented in the following expression
\begin{equation}
    H^2=\frac{-1+\bigg(\sqrt{\frac{5}{4}\alpha \Tilde{V}^3} +\sqrt{1+\frac{5}{4}\alpha \Tilde{V}^3}  \bigg)^{2/3}}{\sqrt{5 \alpha \Tilde{V}}\bigg(\sqrt{\frac{5}{4}\alpha \Tilde{V}^3} +\sqrt{1+\frac{5}{4}\alpha \Tilde{V}^3}  \bigg)^{1/3}}M_p^2,
\end{equation}
here, for simplicity, we introduce a dimensionless variable,
\begin{equation}
    \Tilde{V}=\frac{V}{M_p^4}.
\end{equation}
The right-hand side of the above expression can be expanded in power series of $\Tilde{V}\ll 1$, because the energy scale of inflation ($E_{inf}=V^{1/4}$) is much smaller than the reduced Planck mass. After the series expansion of equation \eqref{h2}, we can find the following solution
\begin{equation}\label{h2f}
H^2\simeq \frac{M_p^2\Tilde{V}}{3}\left(1-\frac{5}{27} \alpha \Tilde{V}^3\right). 
\end{equation}
Now, one can clearly see that the general relativity limit is represented by the first term of the right-hand side, whereas a correction to GR due to the nonminimal coupling between matter and nonmetricity is represented by the second term.

Let us work on the second Friedmann equation \eqref{21}. Now, it can be represented in the following form
\begin{equation}
    \dot{H}\simeq -\frac{\dot{\phi} ^2 \left(\alpha  H^6+M_p^6\right)}{2 M_p^4 \left(\alpha  H^4 \Tilde{V}+M_p^4\right)}.
\end{equation}
Where we have used $\rho+p=\dot{\phi}^2$, here, we have neglected the terms, which include the time derivatives of Hubble parameter $H$ and pressure $p$. 

Using the solution for $H^2$ in the above equation \eqref{h2f} and expanding in the power series of $\Tilde{V}$, we get
\begin{equation}\label{dh}
    \dot{H}\simeq-\frac{\dot{\phi}^2}{2M_p^2}\left(1-\frac{2\alpha}{27} \Tilde{V}^3\right).
\end{equation}
Now, taking time derivative equation \eqref{h2f} with respect to $t$ and using equation \eqref{dh} to eliminate $\dot{H}$, we obtain
\begin{equation}
    \partial_{\phi}{V}\simeq -3 H \dot{\phi}\left(1-\frac{2 \alpha}{3} \Tilde{V}^3\right)\left(1-\frac{20 \alpha}{27} \Tilde{V}^3\right)^{-1},
\end{equation}
and expanding it in the power series of $\Tilde{V}$, we find
\begin{equation}\label{dphi}
    \partial_{\phi}{V}\simeq -3 H \dot{\phi}\left(1+\frac{2 \alpha}{3} \Tilde{V}^3\right).
\end{equation}
Taking the second-time derivative of the above equation \eqref{dphi}, we find
\begin{equation}\label{ddphi}
    \partial^2_{\phi \phi}{V}\simeq 3 H^2 \left(1+\frac{2 \alpha}{3} \Tilde{V}^3\right)\left(-\frac{\dot{H}}{H^2}-\frac{\ddot{\phi}}{H\dot{\phi}}\right),
\end{equation}
where the terms proportional to $\dot{\phi}\Tilde{V}^2$ are neglected. Using equations \eqref{h2f},  \eqref{dh},  \eqref{dphi}, and  \eqref{ddphi}, the quantities $\dot{H}/H^2$ and $\frac{\ddot{\phi}}{H\dot{\phi}}$ can be expressed as
\begin{equation}
    \frac{\dot{H}}{H^2}\simeq \epsilon\left(1-\frac{28}{27}\alpha \Tilde{V}^3\right),
\end{equation}
and,
\begin{equation}
    \frac{\ddot{\phi}}{H\dot{\phi}}\simeq \epsilon\left(1-\frac{28}{27}\alpha \Tilde{V}^3\right)-\eta \left(1-\frac{13}{27}\alpha \Tilde{V}^3\right).
\end{equation}
Now, considering the conditions on the slow-roll inflationary scenario $|\dot{H}|/H^2\ll 1$ and $|\frac{\ddot{\phi}}{H\dot{\phi}}|\ll 1$, we find that
\begin{equation}
    \epsilon \ll 1+\frac{28}{27}\alpha \Tilde{V}^3,
\end{equation}
and,
\begin{equation}
    \eta \ll 1+\frac{13}{27}\alpha \Tilde{V}^3.
\end{equation}

From the above relations, we can verify that $\epsilon\ll 1$ and $\eta\ll 1 $ because $\Tilde{V}=V/M_p^4 \ll 1$ with assumptions $\alpha=\mathcal{O}(1)$. From these results, we conclude that the slow-roll parameter satisfies the inflation conditions $\epsilon\ll 1$ and $\eta\ll 1 $. Further, these parameters are related to the parameters $c_2$ and $c_3$ (presented in \eqref{sc2} and \eqref{sc3}, respectively), which evolve in the de Sitter swampland conjectures through the following relation
\begin{equation}
    c_2^2 < 2\epsilon,
\end{equation}
\begin{equation}
    c_3< |\eta|.
\end{equation}
The above results imply that $c_2^2 \ll 1$ and $c_3 \ll 1$ during a quasi-exponential inflationary scenario. Therefore, the de-Sitter swampland conjectures can not be met for inflation under the framework of nonminimal matter-nonmetricity coupling theories of gravity.

\section{Conclusion}\label{sec5}

In this manuscript, we have examined the swampland conjectures within the framework of nonminimally coupled matter-nonmetricity gravitational theory. In order to do that, we have presumed de Sitter swampland conjectures and the agreement of the slow-roll conditions of inflation. 

The analysis of inflationary regimes under the nonminimal coupled gravity framework leads to some interesting conclusions. It is observed that the inflationary scenario discussed in this work differs from the one in general relativity, and this difference is induced by the choice of $f_2(Q)$ function. Further, we find that the slow-roll conditions are very much controlled by the necessities for the inflationary potential under quite general conditions. Even if the influence of nonminimal coupling theory is determined by the model-free parameter $\alpha$, which could be greater than one and cannot overcome the typical scale of the inflationary potential and its smallness in relation to the Planck scale. We have considered that $\alpha=\mathcal{O}(1)$ for the generality. Combining all of our output and analysis, We infer that the de Sitter swampland criteria cannot be satisfied in the presence of nonminimal matter-nonmetricity gravity coupling. We anticipate that our findings will hold true for any number of inflation fields.\\
Moreover, we are the first to explore this type of investigation in the non-minimally coupled nonmetricity-matter theory of gravity. Therefore, this study opens up many possibilities to explore in the context of this setup in the near future. Some of the widely discussed issues are presented here to look forward. For example, it will be interesting to investigate the recent most debatable $H_0$ tension issue with observational measurements. Also, investigating the profiles of dark matter and dark energy with the swampland conjectures in this theory may shed some light on these important problems in modern cosmology. Further, screening mechanisms such as the chameleon mechanism in this setup would be worth testing. We hope to address some of these issues in the near future.

\section*{Acknowledgements}

S.M. acknowledges Transilvania University of Brasov, Romania, for providing Transilvania postdoctoral research fellowship. The work of KB was supported in part by the JSPS KAKENHI Grant Number JP21K03547.

\end{document}